\documentstyle[a4,12pt]{article}
\def\gtrsim{\mathrel{\mathpalette\vereq>}}

\makeatletter
\def\vereq#1#2{\lower3pt\vbox{\baselineskip1.5pt \lineskip1.5pt
\ialign{$\m@th#1\hfill##\hfil$\crcr#2\crcr\sim\crcr}}}
\makeatother

\begin{document}

\begin{titlepage}

\begin{flushright}
UCB-PTH-01/28 \\
LBNL-48669 \\
\end{flushright}

\vskip 1.5cm

\begin{center}
{\Large \bf  Gauge-Higgs Unification in Higher Dimensions}

\vskip 1.0cm

{\large 
Lawrence Hall, Yasunori Nomura and David Smith
}

\vskip 0.5cm
 {\it Department of Physics, \\ and \\
Theoretical Physics Group, Lawrence Berkeley National Laboratory,\\
University of California, Berkeley, CA 94720}

\vskip 1.0cm

\abstract{
The electroweak Higgs doublets are identified as components of a vector 
multiplet in a higher dimensional supersymmetric field theory. We
construct a minimal model in 6D where the electroweak $SU(2) \otimes U(1)$
gauge group is extended to $SU(3)$, and unified 6D models with the unified
$SU(5)$ gauge symmetry extended to $SU(6)$. In these realistic
theories the extended gauge group is broken by orbifold boundary 
conditions, leaving Higgs doublet zero modes which have Yukawa couplings 
to quarks and leptons on the orbifold fixed points.  In one $SU(6)$ model 
the weak mixing angle receives power law corrections, while in another 
the fixed point structure forbids such corrections.  A 5D model is also 
constructed in which the Higgs doublet contains the fifth component of 
the gauge field. In this case Yukawa couplings are introduced as non-local 
operators involving the Wilson line of this gauge field.}

\end{center}
\end{titlepage}

\section{Introduction}

Precision electroweak data suggest that the weak interactions are
broken by the vacuum expectation value of a scalar field: the Higgs boson. 
Yet, the quadratic divergence in the Higgs mass tells us that the Higgs
should become something other than just a scalar field at energies
not far above the weak scale.  What is the fundamental origin of the 
Higgs boson?  A first step, which we adopt in this paper, is that the 
Higgs is a component of a 4D supersymmetric chiral
multiplet. An economical possibility would be that the Higgs doublet is
identified as the supersymmetric partner of the left-handed lepton
doublet, but this has been hard to implement. This would have made 
significant progress in understanding the origin of the Higgs:
instead of the three types of fields of the standard model
(gauge, chiral matter and Higgs), the supersymmetric theory would
have only two types of fields: vector multiplets and chiral multiplets 
which are chiral under the gauge group. Instead we are driven to the 
minimal supersymmetric theory where a third type of multiplet is
added: a pair of Higgs doublets in chiral multiplets which are 
vector-like under the gauge group.  Even though there are several ways 
to obtain such light vector-like Higgs multiplets, one cannot help but 
feel that the theory would look more elegant without them.

In this paper we study higher dimensional supersymmetric theories in
which there are only two types of fields: a vector multiplet
containing the gauge bosons, and matter multiplets which are chiral 
under the gauge group containing quarks and leptons. 
The gauge group is enlarged beyond that of the standard
model and is broken by compactification on an orbifold, which
nevertheless preserves a single supersymmetry. As expected, the resulting
massless modes are found to include 4D vector multiplets and
chiral multiplets which are chiral under the unbroken gauge group.
We also find that there can be zero modes in chiral multiplets
which are vector-like under the gauge group. These 4D Higgs multiplets
originate from the higher dimensional vector multiplet. In this paper
we identify the Higgs doublets as remnant zero modes of the higher 
dimensional vector multiplet. The vector multiplet transforms in the 
adjoint representation of the gauge group and, for the standard model 
gauge group, this does not contain weak doublets, hence the gauge group
must be enlarged.

A simple implementation of this idea is for the Higgs
doublets to be the higher dimensional components of the gauge fields 
\cite{Manton:1979kb}. This idea does not require supersymmetry --- 
the quadratic divergence of the Higgs mass at low energies is regulated 
by local gauge invariance in the higher dimensions. However, in higher 
dimensional supersymmetric theories, the vector multiplet contains 
scalars of the higher dimensional Lorentz symmetry, allowing alternative
identifications of the Higgs doublets. 

An immediate objection to the Higgs boson originating from a vector 
multiplet is that independent Yukawa couplings of the Higgs to 
matter are forbidden by higher dimensional gauge and supersymmetries. 
Apparently the rich structure of Yukawa couplings of the standard model 
must somehow all arise from the gauge interaction, which would
presumably have to be very complicated. This objection largely disappears 
in the case that the extra dimensional spacetime is an orbifold and that
the orbifold boundary conditions break some of the gauge and 
supersymmetries. At some orbifold fixed points certain gauge 
transformation parameters are constrained to vanish, so that these fixed 
points feel only a restricted gauge symmetry \cite{Hall:2001pg}. 
While the Higgs is a component of the vector multiplet in
the bulk, as far as these fixed points are concerned they are
components of matter type multiplets, and the restricted symmetries
may allow independent Yukawa interactions to be located on these
fixed points.

In this paper we seek to implement this gauge origin for the 
Higgs doublets in simple, realistic effective field theories.
Any such theory involves several choices: the number of extra dimensions,
the number of supersymmetries, the orbifold spacetime and the
gauge group. With a single extra dimension, the 5D vector multiplet
contains a 4D vector multiplet and a 4D chiral adjoint field:
$(V, \Phi)$. The Higgs doublets would lie in $\Phi$, and would
therefore contain $A_5$, the component of the 
gauge field in the fifth dimension. 
In this case, even though the broken gauge transformation
parameters, $\xi(y)$, may vanish on an orbifold fixed point, the
derivatives, $\partial_y \xi$, do not. Thus the Higgs will have an
inhomogeneous transformation under the broken gauge generators,
forbidding independent local Yukawa couplings from appearing at
the fixed point --- the objection of the previous paragraph remains.
This situation is unchanged in 6D with $N=1$ supersymmetry. Furthermore
the 6D $N=1$ vector multiplet has anomalies. We are therefore led to
6D $N=2$ theories. The vector multiplet is anomaly free, and contains
three chiral adjoints $(V,\Phi_i)$. Since there are only two extra
components of the gauge field, $A_{5,6}$, at least one of $\Phi_i$
does not contain any gauge fields and therefore gauge transforms
homogeneously. Higgs doublets arising from such a $\Phi_i$ may have
local Yukawa couplings at fixed points. The three models presented in
sections \ref{sec:su3-su3}, \ref{sec:su6-1} and \ref{sec:t2z4} 
all have such an origin for the Higgs doublets within $N=2$ 
supersymmetry in 6D.

The simplest extension of the standard model gauge group, which gives 
weak doublets in the adjoint representation, is to embed $SU(2) \otimes
U(1)$ into $SU(3)$, and this minimal case is explored in section 
\ref{sec:su3-su3}. Indeed, the only addition to the adjoint is two
weak doublets. This extension does not increase the rank of the gauge
group, and hence the orbifold breaking to the standard model gauge
group is particularly straightforward. In sections \ref{sec:su6-1} and 
\ref{sec:t2z4} we explore extending the $SU(5)$ grand unified theory 
to $SU(6)$. As well as weak doublets, the addition to the adjoint now 
contains color triplets. However, we find that orbifold gauge symmetry 
breaking can remove the unwanted colored triplets, by an extension of 
the $SU(5)$ case \cite{Kawamura:2001ev}.

For a 1D bulk there is a unique orbifold of finite size: $S^1/Z_2$, 
while for 2D there are many possibilities. The Higgs can originate from 
a 6D vector multiplet by symmetry breaking on many 2D orbifolds. In
sections \ref{sec:su3-su3} and \ref{sec:su6-1} we make the simple 
choice of $T^2/(Z_2 \times Z_2')$, which repeats the $S^1/Z_2$ structure 
in both of the extra dimensions. Nevertheless, the fixed point structure
and therefore the nature of the Kaluza-Klein (KK) towers, is very 
sensitive to the orbifold choice. In the $SU(6)$ theory of section 
\ref{sec:su6-1} we find that the KK towers lead to power law running of 
gauge couplings which is not $SU(5)$ invariant. This gives a power law 
correction to the weak mixing angle. In section \ref{sec:t2z4} we 
construct an alternative $SU(6)$ theory, with symmetry breaking on 
$T^2/Z_4$, where such power law corrections are absent.

In section \ref{sec:5d-su6} we return to the case that the Higgs 
doublets contain higher dimensional components of gauge fields. 
Although local Yukawa couplings are forbidden, the Higgs may couple to 
quarks and leptons via non-local interactions involving Wilson lines. 
We do not consider how such non-local interactions may be generated, 
but simply assume that all gauge invariant interactions occur in the 
effective field theory, local or not. In this case we are able to 
construct 5D theories with gauge symmetry broken on $S^1/Z_2$.

Our discussions include further aspects of these models, including
supersymmetry breaking and the location of quarks and leptons. We
discuss the possibility that the third generation resides on an
$SU(5)$ invariant 3 brane, yielding the successful $b/\tau$
mass relation, while the lighter two generations reside on a 4 brane
and therefore have suppressed, non-$SU(5)$ invariant Yukawa couplings.

\section{6D $SU(3)_C \otimes SU(3)_L$ Model on $T^2/(Z_2 \times Z_2')$}
\label{sec:su3-su3}

In this section we present a minimal model which realizes the idea 
that the Higgs fields are components of the gauge supermultiplet 
in higher dimensions.  Unlike previous works \cite{Manton:1979kb},
the Higgs bosons here are not extra dimensional components of the 
gauge field, but rather scalar fields that are supersymmetric 
partners of the gauge field in higher dimensions.

\subsection{Orbifold and Gauge Structure}

We consider a 6D gauge theory with $N=2$ supersymmetry.
The extra dimensions are compactified on a $T^2/(Z_2 \times Z_2')$ 
orbifold with radii $R_5 \sim R_6$.  The $N=2$ supersymmetry 
in 6D corresponds to $N=4$ supersymmetry in 4D, so that only the 
gauge multiplet can be introduced in the bulk.  We take an 
$SU(3)_C \otimes SU(3)_L$ gauge multiplet propagating in the bulk.
This multiplet can be  decomposed under a 4D $N=1$ supersymmetry into a 
vector supermultiplet $V$ and three chiral multiplets $\Sigma_5$, 
$\Sigma_6$ and $\Phi$ in the adjoint representation.
The fifth and sixth components of the gauge field, $A_5$ and $A_6$, 
are contained in the lowest component of $\Sigma_5$ and $\Sigma_6$, 
respectively.

Using the 4D $N=1$ language, the bulk action is written as 
\cite{Arkani-Hamed:2001tb}
\begin{eqnarray}
  S &=& \int d^6 x \Biggl\{
  {\rm Tr} \Biggl[ \int d^2\theta \left( \frac{1}{4 k g^2} 
  {\cal W}^\alpha {\cal W}_\alpha + \frac{1}{k g^2} 
  \left( \Phi \partial_5 \Sigma_6 - \Phi \partial_6 \Sigma_5
  - \frac{1}{\sqrt{2}} \Phi 
  [\Sigma_5, \Sigma_6] \right) \right) + {\rm h.c.} \Biggr] 
\nonumber\\
  && + \int d^4\theta \frac{1}{k g^2} {\rm Tr} \Biggl[ 
  (\sqrt{2} \partial_5 + \Sigma_5^\dagger) e^{-V} 
  (-\sqrt{2} \partial_5 + \Sigma_5) e^{V}
  + (\sqrt{2} \partial_6 + \Sigma_6^\dagger) e^{-V} 
  (-\sqrt{2} \partial_6 + \Sigma_6) e^{V}
\nonumber\\
  && \qquad \qquad \qquad
  + \Phi^\dagger e^{-V} \Phi e^{V} 
  + \partial_5 e^{-V} \partial_5 e^{V}
  + \partial_6 e^{-V} \partial_6 e^{V} \Biggr] \Biggr\},
\label{eq:5daction}
\end{eqnarray}
in the Wess-Zumino gauge.
The $SU(3)_C \otimes SU(3)_L$ gauge transformation is given by
\begin{eqnarray}
  e^V &\rightarrow& e^\Lambda 
    e^V e^{\Lambda^\dagger}, \\
  \Sigma_5 &\rightarrow& e^\Lambda (\Sigma_5 - \sqrt{2} \partial_5) 
    e^{-\Lambda}, \\
  \Sigma_6 &\rightarrow& e^\Lambda (\Sigma_6 - \sqrt{2} \partial_6) 
    e^{-\Lambda}, \\
  \Phi &\rightarrow& e^\Lambda \Phi e^{-\Lambda}.
\end{eqnarray}
Here, we have used a short-handed notation for the 
$SU(3)_C \otimes SU(3)_L$ gauge structure as $\Lambda \equiv 
(\Lambda_C^a T_C^a \oplus \Lambda_L^a T_L^a)$, $V \equiv 
(V_C^a T_C^a \oplus V_L^a T_L^a)$, $\Sigma_5 \equiv (\Sigma_{5,C} 
\oplus \Sigma_{5,L})$, $\Sigma_6 \equiv (\Sigma_{6,C} \oplus 
\Sigma_{6,L})$, and $\Phi \equiv (\Phi_C \oplus \Phi_L)$, 
where $T_C^a$ and $T_L^a$ $(a = 1,\cdots,8)$ are the generators of 
the $SU(3)_C$ and $SU(3)_L$ gauge groups, respectively.
Similarly, the gauge coupling $g$ should also be understood to 
contain two gauge couplings $g_C$ and $g_L$ for $SU(3)_C$ and $SU(3)_L$: 
$1/g^2 \equiv (1/g_C^2 \oplus 1/g_L^2)$.

Note that the $\Sigma_{5,6}$ fields transform non-linearly under the 
gauge transformation, since they contain $A_{5,6}$ as lowest components.
This prevents us from writing down local operators which couple 
$\Sigma_{5,6}$ to matter fields on the orbifold fixed points, 
making it difficult to identify $\Sigma_5$ or $\Sigma_6$ as the 
Higgs field.\footnote{
Matter can couple to $A_{5,6}$ through non-local operators containing 
${\cal P} \exp(\oint (dx^5 A_5 + dx^6 A_6))$, which may be generated 
by integrating out some physics at the compactification scale.
We present a model of this kind in section \ref{sec:5d-su6}.}
Fortunately, the 6D $N=2$ gauge multiplet contains 
an additional adjoint chiral superfield $\Phi$ that does not 
contain components of the higher dimensional gauge bosons and that can
thus couple to the matter fields localized on the fixed point.
We will soon identify components of these superfields as the Higgs 
doublets of the minimal supersymmetric standard model (MSSM). 

We now describe the model, following the notation of 
Ref.~\cite{Barbieri:2001dm}.  The orbifold $T^2/(Z_2 \times Z_2')$ is 
constructed by identifying points of the infinite plane $R^2$ under 
four operations, ${\cal Z}_5: (x^5,x^6) \rightarrow (-x^5,x^6)$, 
${\cal Z}_6: (x^5,x^6) \rightarrow (x^5,-x^6)$, ${\cal T}_5: 
(x^5,x^6) \rightarrow (x^5+2\pi R_5,x^6)$ and ${\cal T}_6: 
(x^5,x^6) \rightarrow (x^5,x^6+2\pi R_6)$.  Here, for simplicity, we 
have taken the two translations ${\cal T}_5$ and ${\cal T}_6$ to be 
in orthogonal directions.
In general, various fields $\varphi(x^5,x^6)$ transform nontrivially 
under these operations as $\varphi(x^5,x^6) \rightarrow 
{\cal Z}_{5,6}[\varphi(x^5,x^6)]$ and $\varphi(x^5,x^6) \rightarrow 
{\cal T}_{5,6}[\varphi(x^5,x^6)]$.  The consistency condition requires 
that these transformations must be symmetries of the bulk action.  
Then, the above identification is made by imposing the conditions 
${\cal Z}_5[\varphi(x^5,x^6)] = {\cal Z}_6[\varphi(x^5,x^6)] = 
{\cal T}_5[\varphi(x^5,x^6)] = {\cal T}_6[\varphi(x^5,x^6)] = 
\varphi(x^5,x^6)$ for all the bulk fields present in the model.

Under the ${\cal Z}_{5,6}$ operations, various fields in a single 
irreducible gauge representation can transform differently.  Thus, 
the gauge symmetry can be broken by identifications under ${\cal Z}_{5,6}$.
We here require $V$ to transform nontrivially under ${\cal Z}_{5,6}$ such 
that only $SU(3)_C \otimes SU(2)_L \otimes U(1)_Y$ components have 
massless modes.  This is accomplished by taking ${\cal Z}_5$ and 
${\cal Z}_6$ identifications as
\begin{eqnarray}
  V(-x^5,x^6) &=& P_Z V(x^5,x^6) P_Z^{-1},
\label{eq:z5-1} \\
  \Sigma_5(-x^5,x^6) &=& - P_Z \Sigma_5(x^5,x^6) P_Z^{-1},
\label{eq:z5-2} \\
  \Sigma_6(-x^5,x^6) &=& P_Z \Sigma_6(x^5,x^6) P_Z^{-1},
\label{eq:z5-3} \\
  \Phi(-x^5,x^6) &=& - P_Z \Phi(x^5,x^6) P_Z^{-1},
\label{eq:z5-4} 
\end{eqnarray}
and
\begin{eqnarray}
  V(x^5,-x^6) &=& P_Z V(x^5,x^6) P_Z^{-1},
\label{eq:z6-1} \\
  \Sigma_5(x^5,-x^6) &=& P_Z \Sigma_5(x^5,x^6) P_Z^{-1},
\label{eq:z6-2} \\
  \Sigma_6(x^5,-x^6) &=& - P_Z \Sigma_6(x^5,x^6) P_Z^{-1},
\label{eq:z6-3} \\
  \Phi(x^5,-x^6) &=& - P_Z \Phi(x^5,x^6) P_Z^{-1},
\label{eq:z6-4} 
\end{eqnarray}
respectively.  Here, $P_Z$ is given by
\begin{equation}
  P_Z = {\rm diag}(1, 1, 1) \oplus {\rm diag}(1, 1, -1).
\end{equation}
Note that various signs appearing in Eqs.~(\ref{eq:z5-1} -- 
\ref{eq:z6-4}) are determined by invariance of the bulk action 
under the ${\cal Z}_{5,6}$ operations.
 
The ${\cal Z}_5$ identification breaks 4D $N=4$ supersymmetry to 
4D $N=2$ supersymmetry (or equivalently, 6D $N=2$ to 6D $N=1$ 
supersymmetry), with $(V,\Sigma_6)$ forming a vector multiplet and 
$(\Sigma_5,\Phi)$ forming a hypermultiplet.  Similarly, the 
${\cal Z}_6$ identification breaks 4D $N=4$ supersymmetry to 
4D $N=2$ supersymmetry, with $(V,\Sigma_5)$ forming a vector multiplet 
and $(\Sigma_6,\Phi)$ forming a hypermultiplet.  This means that the two 
$N=2$ supersymmetries remaining after the ${\cal Z}_5$ and ${\cal Z}_6$ 
operations are different subgroups of the original $N=4$ supersymmetry.
Thus, the combination of ${\cal Z}_5$ and ${\cal Z}_6$ identifications, 
{\it i.e.} the $T^2/(Z_2 \times Z_2')$ compactification, breaks the 
original 6D $N=2$ supersymmetry all the way down to 4D $N=1$ 
supersymmetry.

Under the ${\cal T}_{5,6}$ operations also, various fields in a single 
irreducible gauge representation can transform differently, breaking 
the gauge symmetry.  In the present $SU(3)_C \otimes SU(3)_L$ model, 
we do not introduce this non-trivial transformation in the gauge space 
for the ${\cal T}_{5,6}$ operations.  However, for later use, we write 
${\cal T}_5$ and ${\cal T}_6$ identifications as
\begin{eqnarray}
  V(x^5+2\pi R_5,x^6) &=& P_T V(x^5,x^6) P_T^{-1},
\label{eq:t5-1} \\
  \Sigma_5(x^5+2\pi R_5,x^6) &=& P_T \Sigma_5(x^5,x^6) P_T^{-1},
\label{eq:t5-2} \\
  \Sigma_6(x^5+2\pi R_5,x^6) &=& P_T \Sigma_6(x^5,x^6) P_T^{-1},
\label{eq:t5-3} \\
  \Phi(x^5+2\pi R_5,x^6) &=& P_T \Phi(x^5,x^6) P_T^{-1},
\label{eq:t5-4} 
\end{eqnarray}
and 
\begin{eqnarray}
  V(x^5,x^6+2\pi R_6) &=& P_T V(x^5,x^6) P_T^{-1},
\label{eq:t6-1} \\
  \Sigma_5(x^5,x^6+2\pi R_6) &=& P_T \Sigma_5(x^5,x^6) P_T^{-1},
\label{eq:t6-2} \\
  \Sigma_6(x^5,x^6+2\pi R_6) &=& P_T \Sigma_6(x^5,x^6) P_T^{-1},
\label{eq:t6-3} \\
  \Phi(x^5,x^6+2\pi R_6) &=& P_T \Phi(x^5,x^6) P_T^{-1},
\label{eq:t6-4} 
\end{eqnarray}
respectively, with $P_T$ given by
\begin{equation}
  P_T = {\rm diag}(1, 1, 1) \oplus {\rm diag}(1, 1, 1).
\end{equation}
Thus, Eqs.~(\ref{eq:t5-1} -- \ref{eq:t6-4}) just give periodic boundary 
conditions for all the fields.  Again, no additional signs 
can be introduced in the above transformations due to the requirement 
of invariance of the bulk action under ${\cal T}_{5,6}$.

In general, we could use different $P_Z$ matrices for ${\cal Z}_5$ 
and ${\cal Z}_6$ (Eqs.~(\ref{eq:z5-1} -- \ref{eq:z5-4}) and 
Eqs.~(\ref{eq:z6-1} -- \ref{eq:z6-4})), and different $P_T$ matrices 
for ${\cal T}_5$ and ${\cal T}_6$ (Eqs.~(\ref{eq:t5-1} -- \ref{eq:t5-4}) 
and Eqs.~(\ref{eq:t6-1} -- \ref{eq:t6-4})).  Here we have chosen 
the same $P_Z, P_T$ matrices for the fifth and sixth directions.
This can enhance the symmetry of the system: we have an extra symmetry 
described by $x^5 \leftrightarrow x^6$ and $\Sigma_5 \leftrightarrow 
\Sigma_6$ if $R_5 = R_6$.  This implies that the choice is a natural one.
It may also be important for fixing an unwanted moduli field, $R_5/R_6$ 
and $\theta_T$ (angle between ${\cal T}_5$ and ${\cal T}_6$), 
at the symmetry enhanced point $R_5/R_6 = 1$ and $\theta_T = \pi/2$.
(We comment on the phenomenology of the case $R_5 \gg R_6$ later 
 in this section.)

Having identified all the boundary conditions, let us consider the 
massless bulk fields in the model.  To work this out, 
we consider the transformation properties for the fields under 
${\cal Z}_{5,6}$.  Since massless modes can arise only from fields 
that are even under both $x^5 \rightarrow -x^5$ and 
$x^6 \rightarrow -x^6$, we need only consider the components of 
$V$ and $\Phi$.  Under the parities, the various components transform as
\begin{equation}
  V_C:\: \left( \begin{array}{ccc}
    (+,+) & (+,+) & (+,+) \\ 
    (+,+) & (+,+) & (+,+) \\ 
    (+,+) & (+,+) & (+,+) 
  \end{array} \right),
\qquad
  V_L:\: \left( \begin{array}{cc|c}
    (+,+) & (+,+) & (-,-) \\ 
    (+,+) & (+,+) & (-,-) \\ \hline
    (-,-) & (-,-) & (+,+) 
  \end{array} \right),
\label{eq:V33trans}
\end{equation}
\begin{equation}
  \Phi_C:\: \left( \begin{array}{ccc}
    (-,-) & (-,-) & (-,-) \\ 
    (-,-) & (-,-) & (-,-) \\ 
    (-,-) & (-,-) & (-,-) 
  \end{array} \right),
\qquad
  \Phi_L:\: \left( \begin{array}{cc|c}
    (-,-) & (-,-) & (+,+) \\ 
    (-,-) & (-,-) & (+,+) \\ \hline
    (+,+) & (+,+) & (-,-) 
  \end{array} \right),
\label{eq:phi33trans}
\end{equation}
where the first and second signs represent parities under 
$x^5 \rightarrow -x^5$ and $x^6 \rightarrow -x^6$, respectively.

From Eq.~(\ref{eq:V33trans}) we find that the low-energy gauge group 
is indeed $SU(3)_C \otimes SU(2)_L \otimes U(1)_Y$, which we identify 
as the standard-model gauge group.  In addition to these vector 
multiplets, however, extra massless modes arise from $\Phi$ fields.
The quantum numbers of these extra massless states under 
$SU(3)_C \otimes SU(2)_L \otimes U(1)_Y$ are read off from 
Eqs.~(\ref{eq:phi33trans}) as $({\bf 1},{\bf 2},1/2) \oplus 
({\bf 1},{\bf 2},-1/2)$, which are exactly the correct quantum 
numbers for the two Higgs doublets, $H_U$ and $H_D$, of the MSSM.
(Here, we have normalized the $U(1)_Y$ charges to match convention.)
In the next sub-section, we identify these extra massless states 
as the Higgs doublets and couple them to quarks and leptons 
on the orbifold fixed point.

\subsection{Fixed Points and Quarks and Leptons}

The $T^2/(Z_2 \times Z_2')$ orbifold has four fixed points at 
$(x^5,x^6) = (0,0), (\pi R_5, 0), (0, \pi R_6)$ and $(\pi R_5, \pi R_6)$.
To understand what types of matter fields and interactions can be 
placed on a fixed point, we have to work out the symmetry structures 
of the fixed point \cite{Hall:2001pg, Barbieri:2001dm}.  This can be 
done by investigating the profile of the symmetry transformation 
parameters in the extra dimension.  We find that the gauge 
transformation parameters for $SU(3)_L / (SU(2)_L \otimes U(1)_Y)$ 
vanish on the four fixed points, so that the gauge symmetry on the 
fixed points is $SU(3)_C \otimes SU(2)_L \otimes U(1)_Y$.
These four fixed points are connected by four fixed lines, on which  
the only non-trivial gauge transformations are again those of the 
standard model. As for supersymmetry, three of the four supersymmetry 
transformation parameters vanish on the four fixed points, so that 
the remaining supersymmetry on the fixed points is 4D $N=1$ 
supersymmetry.  (The supersymmetry on the four fixed lines is 
4D $N=2$ supersymmetry.)  Therefore, we find that the 
original bulk symmetry is reduced to 4D $N=1$ supersymmetry and 
$SU(3)_C \otimes SU(2)_L \otimes U(1)_Y$ gauge symmetry on each of 
the four fixed points.  In fact, these four fixed points are completely 
equivalent due to the symmetry of the system.  The matter fields and 
interactions located on the fixed points need (only) respect these 
symmetries.

The 6D $N=2$ supersymmetry prevents us from introducing quarks and 
leptons in the bulk, so that they must be localized on the orbifold 
fixed points or fixed lines. We here introduce quark and lepton chiral 
superfields, $Q, U, D, L$ and $E$, on the $(x^5,x^6) = (0,0)$ fixed point.  
The $N=1$ supersymmetric Yukawa couplings are also introduced on this 
fixed point:
\begin{equation}
  {\mathcal L}_6 \supset \delta(x^5) \delta(x^6) 
    \int d^2\theta \left( \lambda_U Q U H_U + 
    \lambda_D Q D H_D + \lambda_E L E H_D \right),
\label{eq:yukawa}
\end{equation}
where the two Higgs doublets, $H_U$ and $H_D$, are components of the 
$\Phi$ field in the higher dimensional $SU(3)_C \otimes SU(3)_L$ 
gauge multiplet.  With these Yukawa couplings, the theory reduces to 
the MSSM below the compactification scale.

One can also induce small neutrino masses through the see-saw mechanism 
\cite{Seesaw} by introducing right-handed neutrino superfields $N$ 
on the fixed point.  

\subsection{$R$ Symmetry}

The bulk action Eq.~(\ref{eq:5daction}) possesses a $U(1)_R$ symmetry.
This $U(1)_R$ symmetry is extended to the full theory by assigning 
appropriate charges for the quark and lepton superfields.
Here we impose the discrete $Z_4$ subgroup of this $U(1)_R$ symmetry 
on the model.  The charge assignment of this $Z_{4,R}$, which allows 
the Yukawa couplings of Eq.~(\ref{eq:yukawa}) (and the Yukawa couplings 
and Majorana masses for $N$), is given in Table~\ref{tab:z4rsymmetry}.
\begin{table}
\begin{center}
\begin{tabular}{|c|c|cccc|cccccc|} \hline
  & $\theta^\alpha$ & $V$ & $\Sigma_5$ & $\Sigma_6$ 
    & $\Phi$ & $Q$ & $U$ & $D$ & $L$ & $E$ & $N$ 
\\ \hline
  $Z_{4,R}$ & $1$ & $0$ & $0$ & $0$ & $2$ & $-1$ 
    & $1$ & $1$ & $-1$ & $1$ & $1$ 
\\ \hline
\end{tabular}
\end{center}
\caption{$Z_{4,R}$ charge assignment for the 
 $SU(3)_C \otimes SU(3)_L$ model.}
\label{tab:z4rsymmetry}
\end{table}
Imposing the $Z_{4,R}$ symmetry on the theory forbids an unwanted large 
mass term for the Higgs doublets, $[H_U H_D]_{\theta^2}$, on the fixed 
point.  (A mass term for the Higgs fields, $[H_U H_D]_{\theta^2}$, 
of the order of the electroweak scale is generated through the 
$Z_{4,R}$ breaking effect after supersymmetry is broken.)
This symmetry contains the $R$-parity of the MSSM and thus
forbids dangerous operators such as $[LH_U]_{\theta^2}$,
$[QDL]_{\theta^2}$, $[UDD]_{\theta^2}$ and $[LLE]_{\theta^2}$; it also
forbids the $d=5$ proton decay operators 
$[QQQL]_{\theta^2}$ and $[UUDE]_{\theta^2}$.  

\subsection{Supersymmetry Breaking}

In the present model the gauge and Higgs multiplets propagate in the 
bulk and the matter fields are localized on the fixed point at 
$(x^5,x^6) = (0,0)$.  This provides a natural setting \cite{Hall:2001pg} 
for gaugino mediated supersymmetry breaking \cite{Kaplan:2000ac}.
Supersymmetry is broken by the $F$-component expectation value, 
$F_S$, of a field $S$ on either of the three fixed points $(x^5,x^6) 
= (\pi R_5, 0), (0, \pi R_6)$ or $(\pi R_5, \pi R_6)$, and it is 
directly transmitted to the gauge and Higgs multiplets 
through the operators
\begin{equation}
  {\mathcal L}_6 = \delta(x^5-x_f^5) \delta(x^6-x_f^6) 
    \left[ \int d^2\theta S {\cal W}_i^{\alpha} {\cal W}_{i\alpha} +  
    \int d^4\theta (S^\dagger H_U H_D + 
    S^\dagger S H_U H_D) + {\rm h.c.} \right].
\label{eq:gm}
\end{equation} 
Here, $(x_f^5, x_f^6)$ is the coordinate of the fixed point where 
$S$ field is localized, and we have omitted coefficients of order unity 
in units of the fundamental scale.  Note that $S$ has a vanishing 
$Z_{4,R}$ charge, so that $F_S$ breaks the $Z_{4,R}$ symmetry.  
The interactions of Eq.~(\ref{eq:gm}) generate gaugino 
masses as well as the $\mu$ and $\mu B$ parameters, while 
the squarks and sleptons obtain masses through radiative corrections 
so that the supersymmetric flavor problem is naturally solved.
Since only the $SU(3)_C \otimes SU(2)_L \otimes U(1)_Y$ gauge symmetry 
is preserved on the fixed point, the masses for the three gauginos 
can take different values.

\subsection{Gauge Couplings and Compactification Scale}

The $SU(3)_L$ unification in our model suggests that $g_1^{3 \otimes 3} 
= g_2$ should be satisfied at the cutoff $M_*$.  
Here, $g_1^{3 \otimes 3}$ is the correctly normalized 
hypercharge gauge coupling when the hypercharge operator is identified 
as the appropriate $SU(3)_C \otimes SU(3)_L$ generator $T_Y$, 
satisfying ${\rm Tr} (T_Y)^2=1/2$.  This boundary condition may instead 
be expressed in terms of $g_Y$, the conventionally normalized 
hypercharge coupling of the standard model.  The 
$SU(3)_C \otimes SU(3)_L$ hypercharge generator is 
$T_Y = (1/2\sqrt{3})[{\rm diag}(0,0,0) \oplus {\rm diag}(1,1,-2)]$, 
leading to hypercharge assignments for the Higgs doublets equal to 
$\pm \sqrt{3}/2$.  These assignments are a factor $\sqrt{3}$ larger than 
the standard model ones, implying that $g_Y=\sqrt{3} g_1^{3 \otimes 3}$, 
so that the correct boundary condition is $g_Y(M_*) = \sqrt{3}
g_2(M_*)$.

If only massless zero modes contributed to the running of $g_Y$ and
$g_2$, this boundary condition would require $M_*$ to be
well above the Planck scale for low energy data to be reproduced.  
However, at scales above the
compactification scale $M_c$, $g_Y$ and $g_2$ undergo power-law
running \cite{Dienes:1998vh}, allowing $M_*$ to be lowered.\footnote{
As will be discussed in sub-section \ref{subsec:su6running}, the 4D 
$N=4$ supersymmetry of the bulk is broken to $N=2$ on 5D fixed lines of 
the orbifold, leading to a linear (rather than quadratic) evolution of 
the gauge couplings above $M_c$.}
By calculating the one-loop contributions to the running from the KK 
excitations of the 6D $N=2$ vector multiplet we find that 
$\sqrt{3}g_2$ and $g_1$ unify beneath $M_{Pl}$ provided that 
$M_*/M_c \gtrsim 40$ holds.  Actually, this perturbative calculation 
is not trustworthy because the classical scaling of the gauge couplings 
makes the theory strongly coupled.  At the scale $\mu>M_c$, the 
appropriate loop expansion parameter is $(\alpha / 4 \pi) (\mu/M_c)^2$.  
Taking $\mu=40 M_c$, this parameter is larger than unity even 
for $\alpha=0.01$.  We are thus forced to conclude that the $SU(3)_L$
unification in this model is achieved in a non-perturbative regime.

\subsection{Asymmetric Extra Dimension}

So far, we have been considering $R_5 \sim R_6$.
In this sub-section we comment on the phenomenology of the case 
where there is a (mild) hierarchy between $R_5$ and $R_6$.
We consider the case $R_5 \gg R_6$ without a loss of generality.
In this case, between the two energy scales $R_5^{-1}$ and $R_6^{-1}$, 
the theory appears as 5D $SU(3)_C \otimes SU(2)_L 
\otimes U(1)_Y$ gauge theory with the gauge multiplets and 
two Higgs doublet hypermultiplets in the bulk.
Therefore, if $R_5$ is as low as TeV, there is the possibility that 
the theories discussed in Refs.~\cite{Pomarol:1998sd, 
Arkani-Hamed:2001mi} are low energy effective theories of the 
present $SU(3)_C \otimes SU(3)_L$ model.  Then, if quarks and leptons 
are localized on the fixed point $(x^5, x^6) = (0, 0)$, the lower bound 
on the scale $R_5^{-1}$ comes from the production of single gauge KK 
modes with nonzero KK momentum in the fifth dimension and the 
generation of four zero-mode fermion operators \cite{Nath:1999fs}, 
requiring $R_5^{-1}$ to be larger than a few TeV.

However, instead of putting quarks and leptons on the fixed point, 
we could put them on the 5D fixed line $x^6 = 0$, since the gauge 
symmetry preserved on the fixed line is only that of the standard model.
Although the fixed line preserves 4D $N=2$ supersymmetry and quarks 
and leptons have to be introduced as hypermultiplets, the zero mode 
matter content is precisely that of the MSSM due to the orbifold 
operation ${\cal Z}_5$.  In this case, the bound on $R_5^{-1}$ is 
significantly weaker, since the effects giving a strong bound are 
absent due to the conservation of the KK momentum in the fifth 
dimension \cite{Barbieri:2001vh, Appelquist:2001nn}.

\section{6D $SU(6)$ Unified Model on $T^2/(Z_2 \times Z_2')$}
\label{sec:su6-1}

In this section we construct a model that realizes the idea 
of Higgs fields as components of higher dimensional gauge 
supermultiplet, in which all the standard model gauge groups are 
unified into a single gauge group.

\subsection{Orbifold and Gauge Structure}

We consider a 6D $N=2$ supersymmetric gauge theory as in the previous 
section.  The extra dimensions are compactified on the 
$T^2/(Z_2 \times Z_2')$ orbifold with radii $R_5 \sim R_6 \sim M_U^{-1}$,
where $M_U \simeq 2 \times 10^{16}~{\rm GeV}$ is the conventional 
grand unification scale.
We here set two radii equal, $R \equiv R_5 = R_6$, for simplicity.
The gauge group is taken to be $SU(6)$, so that the only bulk field is 
the $SU(6)$ gauge multiplet, $(V, \Sigma_5, \Sigma_6, \Phi)$.
The orbifold boundary conditions are given by Eqs.~(\ref{eq:z5-1} -- 
\ref{eq:z6-4}) and Eqs.~(\ref{eq:t5-1} -- \ref{eq:t6-4}) with 
$P_Z$ and $P_T$ given by
\begin{eqnarray}
  P_Z &=& {\rm diag}(1, 1, 1, 1, 1, -1),
\\
  P_T &=& {\rm diag}(1, 1, 1, -1, -1, -1).
\end{eqnarray}
This breaks the $SU(6)$ gauge group to $SU(3)_C \otimes SU(2)_L 
\otimes U(1)_Y \otimes U(1)_X$ at low energies.

To identify the massless fields, we consider the transformation 
properties of $V$ and $\Phi$ fields under ${\cal Z}_{5,6}$ and 
${\cal T}_{5,6}$.  They are written as
\begin{equation}
  V:\: \left( \begin{array}{ccc|cc|c}
    (+,+) & (+,+) & (+,+) & (+,-) & (+,-) & (-,-) \\ 
    (+,+) & (+,+) & (+,+) & (+,-) & (+,-) & (-,-) \\ 
    (+,+) & (+,+) & (+,+) & (+,-) & (+,-) & (-,-) \\ \hline
    (+,-) & (+,-) & (+,-) & (+,+) & (+,+) & (-,+) \\ 
    (+,-) & (+,-) & (+,-) & (+,+) & (+,+) & (-,+) \\ \hline
    (-,-) & (-,-) & (-,-) & (-,+) & (-,+) & (+,+)
  \end{array} \right),
\label{eq:V6trans}
\end{equation}
\begin{equation}
  \Phi:\: \left( \begin{array}{ccc|cc|c}
    (-,+) & (-,+) & (-,+) & (-,-) & (-,-) & (+,-) \\ 
    (-,+) & (-,+) & (-,+) & (-,-) & (-,-) & (+,-) \\ 
    (-,+) & (-,+) & (-,+) & (-,-) & (-,-) & (+,-) \\ \hline
    (-,-) & (-,-) & (-,-) & (-,+) & (-,+) & (+,+) \\ 
    (-,-) & (-,-) & (-,-) & (-,+) & (-,+) & (+,+) \\ \hline
    (+,-) & (+,-) & (+,-) & (+,+) & (+,+) & (-,+)
  \end{array} \right),
\label{eq:phi6trans}
\end{equation}
where the first and second signs represent parities under 
${\cal Z}_{5,6}$ and ${\cal T}_{5,6}$, respectively.
We see from Eq.~(\ref{eq:V6trans}) that the massless vector multiplets 
are those of $SU(3)_C \otimes SU(2)_L \otimes U(1)_Y \otimes U(1)_X$.
In addition, we have massless chiral superfields coming from $\Phi$ 
whose quantum numbers are given by $({\bf 1},{\bf 2},1/2,-2) \oplus 
({\bf 1},{\bf 2},-1/2,2)$.  (We have normalized the $U(1)_{Y,X}$ 
charges to match the convention.)  Since these quantum numbers are 
exactly those for the two Higgs doublets of the MSSM, we identify 
these massless states to be the Higgs fields.  The Yukawa couplings 
to quarks and leptons are discussed in the next sub-section.

We here comment on the uniqueness of obtaining massless Higgs doublets.
Note that we could have chosen $P_T = {\rm diag}(1, 1, -1, -1, -1, -1)$
for the purpose of breaking the gauge group to 
$SU(3)_C \otimes SU(2)_L \otimes U(1)_Y \otimes U(1)_X$.
In that case, however, the massless states coming from $\Phi$ are 
triplet Higgs fields, $({\bf 3},{\bf 1},-1/3,-2) \oplus 
({\bf 3}^*,{\bf 2},1/3,2)$, instead of doublets.  

\subsection{Fixed Points and Quarks and Leptons}

The structure of the fixed points can be worked out by considering the 
profiles of symmetry transformation parameters in the extra dimensions.
On each of the four fixed points of the $T^2/(Z_2 \times Z_2')$ orbifold, 
the remaining supersymmetry and gauge symmetry is given in 
Table~\ref{tab:fixedpoints} ---
\begin{table}
\begin{center}
\begin{tabular}{|c|c|c|} \hline
  $(x^5, x^6)$     & 4D supersymmetry & gauge symmetry 
\\ \hline
  $(0,   0)$       & $N=1$ & $SU(5) \otimes U(1)_X$ \\
  $(\pi R, 0)$     & $N=1$ & $SU(3)_C \otimes SU(2)_L \otimes 
                              U(1)_Y \otimes U(1)_X$ \\
  $(0, \pi R)$     & $N=1$ & $SU(3)_C \otimes SU(2)_L \otimes 
                              U(1)_Y \otimes U(1)_X$ \\
  $(\pi R, \pi R)$ & $N=1$ & $SU(4)_{\tilde{C}} \otimes 
                              SU(2)_L \otimes U(1)_{\tilde{X}}$ 
\\ \hline
\end{tabular}
\end{center}
\caption{Supersymmetry and gauge symmetry on each of the four fixed points.}
\label{tab:fixedpoints} 
\end{table}
matter multiplets and interactions placed on the fixed 
points must respect these symmetries. The fixed points are connected 
by fixed lines on which the gauge symmetries are $SU(5) \otimes U(1)_X$ 
for $x_5 = 0$ and $x_6 = 0$, and $SU(4)_{\tilde{C}} \otimes SU(2)_L 
\otimes U(1)_{\tilde{X}}$ for $x_5 = \pi R$ and $x_6 = \pi R$.

We first consider putting quark and lepton superfields on the 
$(x^5,x^6) = (0,0)$ fixed point.  Since the gauge symmetry preserved 
on this fixed point is $SU(5) \otimes U(1)_X$, we have to introduce 
matter and interactions respecting this symmetry.  This $SU(5)$ 
contains unbroken $SU(3)_C \otimes SU(2)_L \otimes U(1)_Y$ as in the 
conventional way \cite{Georgi:1974sy}.  Thus, we introduce 
three generations of quarks and leptons, $3 \times [T({\bf 10}, 1),$ 
$\bar{F}({\bf 5}^*, -3),$ $N({\bf 1}, 5)]$, and couple them to the 
Higgs fields as
\begin{equation}
  {\mathcal L}_6 = \delta(x^5) \delta(x^6) 
    \int d^2\theta (y_T T T H + y_F T \bar{F} \bar{H} + y_N \bar{F} N H) 
    + {\rm h.c.}
\end{equation} 
Here, $H({\bf 5}, -2)$ and $\bar{H}({\bf 5}^*, 2)$ are components of 
the $\Phi$ field whose wavefunctions are nonvanishing at 
$(x^5,x^6) = (0,0)$.  Below the compactification scale, these 
interactions give the usual MSSM Yukawa couplings plus neutrino Yukawa 
couplings, since the only massless fields in $H$ and $\bar{H}$ are 
the doublet components.  These couplings, however, precisely 
respect $SU(5)$ relations leading to unwanted predictions such as 
$m_s/m_d = m_\mu/m_e$. These relations can be avoided by mass mixing with
heavy matter propagating on the 5D fixed lines \cite{Hall:2001pg}. 
Below we follow an alternative choice with quarks and 
leptons on another fixed point, even though this loses 
some understanding of the fermion quantum numbers.  
Further alternative models are discussed 
in sub-sections \ref{subsec:5dfermion} and \ref{subsec:hybrid}.

We locate quark and lepton chiral superfields on the fixed 
point where only $SU(3)_C \otimes SU(2)_L \otimes U(1)_Y \otimes U(1)_X$ 
gauge symmetry is preserved.  Since two fixed points at $(x^5, x^6) 
= (\pi R, 0)$ and $(0, \pi R)$ are equivalent, we put them on the 
$(x^5, x^6) = (\pi R, 0)$ fixed point without a loss of generality.
We introduce three generations of quarks and leptons, 
$3 \times [Q({\bf 3}, {\bf 2}, 1/6, 1),$ $U({\bf 3}^*, {\bf 1}, -2/3, 1),$ 
$D({\bf 3}^*, {\bf 1}, 1/3, -3),$ $L({\bf 1}, {\bf 2}, -1/2, -3),$ 
$E({\bf 1}, {\bf 1}, 1, 1),$ $N({\bf 1}, {\bf 1}, 0, 5)]$, and couple 
them to the Higgs fields as
\begin{equation}
  {\mathcal L}_6 = \delta(x^5) \delta(x^6-\pi R) 
    \int d^2\theta (y_U Q U H_U + y_D Q D H_D + y_E L E H_D + y_N L N H_U)
    + {\rm h.c.}
\label{eq:su6-yukawa}
\end{equation} 
Here, $H_U({\bf 1}, {\bf 2}, 1/2, -2)$ and 
$H_D({\bf 1}, {\bf 2}, -1/2, 2)$ are massless Higgs doublets coming 
from the $\Phi$ field, whose wavefunctions are nonvanishing at 
$(x^5,x^6) = (\pi R, 0)$.  Since these Yukawa couplings need only
respect $SU(3)_C \otimes SU(2)_L \otimes U(1)_Y \otimes U(1)_X$ gauge 
symmetry, there are no unwanted $SU(5)$ fermion mass relations.
Moreover, $d=5$ proton decay due to triplet Higgs exchange 
\cite{Sakai:1982pk} is absent, since there is no coupling of triplet 
Higgs fields to quarks and leptons.\footnote{
We can write down couplings of triplet Higgs fields to quarks and 
leptons using derivatives of the extra dimensional coordinates.  
Even then, however, the mechanism of Ref.~\cite{Hall:2001pg} 
ensures that the $d=5$ proton decay is not caused by the exchange of 
the triplet Higgs fields.}
Similarly, $d=6$ proton decay induced by the exchange of an $X$ gauge 
boson is also absent, since the wavefunction of the $X$ gauge boson 
vanishes on this fixed point \cite{Dienes:1998vh}.\footnote{
There could be operators which couple to the $X$ gauge boson to 
quarks and leptons through the derivative of the extra dimensional 
coordinates, but these operators are suppressed by the volume of the 
extra dimensions and thus expected to be small.}
The $U(1)_X$ symmetry breaking and neutrino masses are discussed 
in the next sub-section.

\subsection{$U(1)_X$ Symmetry Breaking}
\label{subsec:su6u1x}

We have seen that, after the orbifolding, the $SU(6)$ bulk gauge 
multiplet provides massless modes of 4D $N=1$ vector superfields of the 
$SU(3)_C \otimes SU(2)_L \otimes U(1)_Y \otimes U(1)_X$ gauge group and 
two Higgs chiral superfields $H_U$ and $H_D$.
To recover the MSSM at low energies, we have to break the $U(1)_X$ 
gauge symmetry.  This $U(1)_X$ symmetry is the extra $U(1)$ symmetry 
in the usual $SO(10)$ grand unified theory, $U(1)_X = SO(10)/SU(5)$, 
as far as the quantum numbers for the quarks, leptons and Higgs fields 
are concerned.  We here break it with the usual Higgs mechanism 
by introducing chiral superfields $X({\bf 1}, {\bf 1}, 0, 10)$ and 
$\bar{X}({\bf 1}, {\bf 1}, 0, -10)$ on the $(x^5,x^6) = (\pi R, 0)$ 
fixed point.  We consider the following superpotential 
\begin{equation}
  {\mathcal L}_6 = \delta(x^5) \delta(x^6-\pi R) 
    \int d^2\theta \left\{ Y (X \bar{X} - M_X^2) + \bar{X} N^2 
    \right\} + {\rm h.c.},
\end{equation} 
where $Y$ is a singlet superfield.  This superpotential forces 
$X$ and $\bar{X}$ to have vacuum expectation values 
$\langle X \rangle = \langle \bar{X} \rangle = M_X$.
It also gives Majorana masses for the right-handed neutrinos of order 
$M_X$, generating small neutrino masses though the see-saw mechanism 
\cite{Seesaw}.  Motivated by the observation of atmospheric neutrino 
oscillation \cite{Fukuda:1998mi}, we take $M_X \sim 10^{14}~{\rm GeV}$.
An interesting point of breaking $U(1)_X$ by $\langle X \rangle 
= \langle \bar{X} \rangle \neq 0$ is that it leaves an
unbroken $Z_2$ discrete gauge symmetry at low energies, which is 
precisely the matter parity in the MSSM.  Therefore, unwanted operators 
such as $[LH_U]_{\theta^2}$, $[QDL]_{\theta^2}$, $[UDD]_{\theta^2}$ and 
$[LLE]_{\theta^2}$ are never generated even by quantum gravitational 
effects \cite{Krauss:1989zc}.

\subsection{$R$ symmetry}
\label{subsec:su6-R}

To make the model fully realistic, we have to forbid dangerous 
tree-level $d=5$ proton decay operators, $[QQQL]_{\theta^2}$ and 
$[UUDE]_{\theta^2}$, as well as the tree-level Higgs mass term, 
$[H_U H_D]_{\theta^2}$.  This can be done by imposing the discrete 
$Z_{4,R}$ symmetry on the model, whose charge assignment is 
given in Table~\ref{tab:z2rsymmetry}.
\begin{table}
\begin{center}
\begin{tabular}{|c|c|cccc|cccccc|ccc|} \hline
  & $\theta^\alpha$ & $V$ & $\Sigma_5$ & $\Sigma_6$ & $\Phi$ 
    & $Q$ & $U$ & $D$ & $L$ & $E$ & $N$ & $X$ & $\bar{X}$ & $Y$ 
\\ \hline
  $Z_{4,R}$ & $1$ & $0$ & $0$ & $0$ & $2$ & $0$ 
    & $0$ & $0$ & $0$ & $0$ & $0$ & $2$ & $2$ & $2$ 
\\ \hline
\end{tabular}
\end{center}
\caption{$Z_{4,R}$ charge assignment for the $SU(6)$ model.}
\label{tab:z2rsymmetry} 
\end{table}
This $Z_{4,R}$ could be gauged if we employ the Green-Schwarz mechanism 
\cite{Green:1984sg} to cancel anomalies \cite{Ibanez:1991hv}.
The expectation values $\langle X \rangle = \langle \bar{X} \rangle 
\neq 0$ break both $Z_{4,R}$ and $U(1)_X$ symmetries, but it leaves 
another unbroken discrete $Z'_{4,R}$ symmetry that is a linear combination 
of $Z_{4,R}$ and $U(1)_X$: $Z'_{4,R} = Z_{4,R} + (1/5) U(1)_X$.
(To make all charges integer, we have to take a linear combination, 
 $Z_{4,R} + (1/5) U(1)_X + (24/5) U(1)_Y$.)
This $Z'_{4,R}$ symmetry is sufficient to forbid the above unwanted 
operators, and thus no large $\mu$ term is generated 
by the $Z_{4,R}$-$U(1)_X$ breaking.  A $\mu$ term of the order of 
the weak scale is generated though the $Z'_{4,R}$ 
breaking effect after supersymmetry is broken.

\subsection{Supersymmetry Breaking}
\label{subsec:su6susybreak}

Since we have put matter on the $(x^5,x^6) = (\pi R, 0)$ fixed point, 
supersymmetry breaking must happen either on the $(x^5,x^6) = (0,0), 
(0, \pi R)$ or $(\pi R, \pi R)$ brane for gaugino mediation 
to work.  Supersymmetry is broken by the $F$-component expectation 
value, $F_S$, of a field $S$ localized on the fixed point.  
Supersymmetry breaking effects are transmitted to the gauge and Higgs 
multiplets through the operators given in Eq.~(\ref{eq:gm}).
Since the $Z'_{4,R}$ charge of $S$ is zero, $F_S$ breaks the 
$Z'_{4,R}$ symmetry, generating gaugino masses and the $\mu$ and 
$\mu B$ parameters.

On the three fixed points, $(x^5,x^6) = (0,0), (0, \pi R)$ and 
$(\pi R, \pi R)$, the ``unbroken'' gauge groups are different 
as was shown in Table~\ref{tab:fixedpoints}.  Therefore, depending on 
where supersymmetry breaking occurs, we obtain different relations 
for the gaugino masses.  In the case that supersymmetry is broken at 
$(x^5,x^6) = (0,0)$ ($S$ is located on $(x^5,x^6) = (0,0)$), 
we obtain the gaugino mass relations $m_{SU(3)_C} = m_{SU(2)_L} 
= m_{U(1)_Y}$ at the compactification scale, since the interactions 
in Eq.~(\ref{eq:gm}) must respect the $SU(5) \otimes U(1)_X$ gauge 
symmetry remaining on this fixed point.  
On the other hand, if $S$ is located on the 
$(x^5,x^6) = (0,\pi R)$ fixed point, the four gaugino masses, 
$m_{SU(3)_C}, m_{SU(2)_L}, m_{U(1)_Y}$ and $m_{U(1)_X}$, can take 
arbitrary values, since the interactions in Eq.~(\ref{eq:gm}) need 
only respect $SU(3)_C \otimes SU(2)_L \otimes U(1)_Y \otimes U(1)_X$
gauge symmetry.  Finally, if $S$ is on $(x^5,x^6) = (\pi R,\pi R)$, 
we obtain a relation among $m_{SU(3)_C}, m_{U(1)_Y}$ and $m_{U(1)_X}$, 
but it is irrelevant for low energy phenomenology.

\subsection{Gauge Couplings and Compactification Scale}
\label{subsec:su6running}

In the present $SU(6)$ model, the $SU(3)_C$, $SU(2)_L$ and $U(1)_Y$ 
gauge groups are unified into $SU(5)$ as in the conventional way.
Therefore, the compactification scale is given by $1/R = M_U$ in 
the zero-th order approximation.  There are, however, two types of 
corrections to this naive identification \cite{Hall:2001pg, 
Nomura:2001mf}.

First, we can write down tree-level gauge kinetic terms that do not 
respect full $SU(6)$ symmetry on subspaces of the 6D spacetime.
Specifically, we can write 5D gauge kinetic terms respecting only 
$SU(5) \otimes U(1)_X$ gauge symmetry on the $(4+1)$-dimensional spaces 
$x^5 = 0$ and $x^6 = 0$.  Similarly, 5D gauge kinetic terms 
respecting only $SU(4)_{\tilde{C}} \otimes SU(2)_L \otimes 
U(1)_{\tilde{X}}$ can be written on the $(4+1)$-dimensional spaces 
$x^5 = \pi R$ and $x^6 = \pi R$.  Finally, 4D gauge kinetic terms are
also introduced on the four fixed points, which need only respect 
gauge symmetries specified in Table~\ref{tab:fixedpoints}.
However, the corrections from these operators are generically suppressed 
by the volume of the extra dimension(s), so that we will neglect 
these contributions in the following analysis.

The second correction originates from the running of the gauge 
couplings above the compactification scale due to KK modes.  
Since the present model is a 6D theory, the zero-mode gauge couplings 
$g_{0i}$ at the compactification scale $M_c$ ($\equiv 1/R$) receive 
power-law corrections as \cite{Dienes:1998vh}
\begin{eqnarray}
  \frac{1}{g_{0i}^2(M_c)} \simeq 
    \frac{1}{g_0^2(M_*)} 
    - \frac{b}{8 \pi^2} ((M_* R)^2 - 1)
    - \frac{b'_i}{8 \pi^2} (M_* R - 1)
    + \frac{b''_i}{8 \pi^2} \ln(M_* R),
\label{eq:powerlaw}
\end{eqnarray}
where $b, b'_i$ and $b''_i$ are constants of $O(1)$ and $M_*$ the 
cutoff scale of the theory.  In 6D picture, 
the last three terms correspond to 6D, 5D and 4D gauge kinetic terms 
generated by loop effects in the 6D bulk, on the 5D subspaces and 
the 4D fixed points, respectively.  An interesting point is that 
the present model possesses 6D $N=2$ supersymmetry in the bulk, 
so that the term quadratically sensitive to the cutoff does not 
appear, $b=0$.  The other two terms give the correction to the 
relation between $M_c$ and $M_U$ and to the standard 4D
supersymmetric $SU(5)$ prediction of $\sin^2\theta_w$.

To estimate the threshold correction coming from this second 
source, we consider the one-loop renormalization group equations for the
three gauge couplings.  Assuming that the couplings take a unified value 
$g_*$ at $M_*$, they take the following form:
\begin{eqnarray}
  \alpha_i^{-1}(m_Z) &=& \alpha_*^{-1}(M_*)
    + \frac{1}{2\pi} \Biggl\{ \alpha_i \ln\frac{m_{\rm SUSY}}{m_Z} 
    + \beta_i \ln\frac{M_*}{m_Z} 
    + \gamma_i \sum_{n=0}^{N_l} \ln\frac{M_*}{(n+1)M_c} \Biggr\}, 
\label{eq:rge}
\end{eqnarray}
where $(\alpha_1, \alpha_2, \alpha_3) = (-5/2, -25/6, -4)$, 
$(\beta_1, \beta_2, \beta_3) = (33/5, 1, -3)$ and 
$(\gamma_1, \gamma_2, \gamma_3) = (12/5, -4, -12)$.
Here, we have assumed a common mass $m_{\rm SUSY}$ for the superparticles 
for simplicity, and the sum on $n$ includes all KK modes 
below $M_*$, so that $(N_l+1)M_c \leq M_*$.  As expected, the term 
which involves the double sum of the KK towers does not appear, but
there is still non-universal linear running
of the gauge couplings above the compactification scale, as indicated
by the presence of the single sum.  This power-law contribution yields
a large threshold correction that could spoil the successful 
prediction of $\sin^2\theta_w$, if $N_l$ is taken too large. 
For $N_l=3$, we estimate the threshold correction to the
value of $\sin^2 \theta_w$ to be $\sim (2-3)$\%,  
while for $N_l=10$, the correction is $\sim 10 \%$.  Consistency with
low-energy data requires some degree of cancellation between threshold
corrections coming from unknown cutoff-scale physics 
and this correction arising from the vector multiplet KK modes.
In section \ref{sec:t2z4}, we describe a slightly different $SU(6)$ 
model in which the vector multiplet KK excitations give a vanishing 
contribution to the gauge coupling running so that the successful 
prediction of $\sin^2 \theta_w$ is preserved.

\subsection{Quarks and Leptons in 5D subspace}
\label{subsec:5dfermion}

So far, we have been considering a model with quarks and leptons  
localized on a 4D fixed point.  In this sub-section, we present an 
alternative model in which they  propagate in a 5D subspace of the 
original 6D spacetime.

First, we note that the $T^2/(Z_2 \times Z_2')$ orbifold is equivalent to 
an $(S^1/Z_2)^2$ orbifold where each of the fifth and sixth dimensions is 
compactified on $S^1/Z_2$.  Therefore, we can imagine the situation 
where matter is localized on the fixed point of the $S^1/Z_2$ 
in the $x^5$ direction but is freely propagating in the $x^6$ direction.
This means that $T^2/(Z_2 \times Z_2')$ has four $(4+1)$-dimensional 
subspaces where matter and/or interactions can be placed.  
On each subspace, which we call a fixed line, the remaining supersymmetry 
and gauge symmetry are given in Table~\ref{tab:fixedlines}.
\begin{table}
\begin{center}
\begin{tabular}{|c|c|c|} \hline
  fixed lines      & 4D supersymmetry & gauge symmetry 
\\ \hline
  $x^5 = 0$        & $N=2$ & $SU(5) \otimes U(1)_X$ \\
  $x^6 = 0$        & $N=2$ & $SU(5) \otimes U(1)_X$ \\
  $x^5 = \pi R$    & $N=2$ & $SU(4)_{\tilde{C}} \otimes 
                              SU(2)_L \otimes U(1)_{\tilde{X}}$ \\
  $x^6 = \pi R$    & $N=2$ & $SU(4)_{\tilde{C}} \otimes 
                              SU(2)_L \otimes U(1)_{\tilde{X}}$ 
\\ \hline
\end{tabular}
\end{center}
\caption{Supersymmetry and gauge symmetry on each of the four fixed lines.}
\label{tab:fixedlines} 
\end{table}

We here locate quarks and leptons on the fixed line where 
$SU(5) \otimes U(1)_X$ gauge symmetry is preserved.  Since the two 
fixed lines $x^5 = 0$ and $x^6 = 0$ are equivalent, we choose to 
put quarks and leptons on the $x^6 = 0$ fixed line without a loss 
of generality.  Then, we have to introduce quarks and leptons 
in the form of hypermultiplets, since 4D $N=2$ supersymmetry 
(5D $N=1$ supersymmetry) is preserved on the fixed line.  
We find that introducing only three hypermultiplets 
${\cal T}({\bf 10}, 1)$, ${\cal F}({\bf 5}^*, -3)$ and 
${\cal N}({\bf 1}, 5)$ for each generation does not work due to 
an automatic ``double-triplet splitting'' mechanism caused by 
non-trivial matrix $P_T$.  Rather, we have to introduce at least 
five hypermultiplets, ${\cal T}({\bf 10}, 1)$, ${\cal T}'({\bf 10}, 1)$, 
${\cal F}({\bf 5}^*, -3)$, ${\cal F}'({\bf 5}^*, -3)$ and 
${\cal N}({\bf 1}, 5)$, to obtain the correct low energy matter content 
\cite{Hall:2001pg, Hebecker:2001wq, Barbieri:2001yz}. 

We explicitly show the boundary conditions for the matter fields 
located on the fixed line.  Each hypermultiplet ${\cal M}$ 
(${\cal M} = {\cal T}, {\cal T}', {\cal F}, {\cal F}', {\cal N}$) 
is decomposed into two chiral superfields $M^{(+)}$ and $M^{(-)}$ under 
4D $N=1$ supersymmetry.  Then, the boundary conditions are written as
\begin{equation}
  M^{(\pm)}(-x^5,x^6) = \pm M^{(\pm)}(x^5,x^6),
\end{equation}
and
\begin{eqnarray}
  T^{(\pm)}(x^5+2\pi R_5,x^6) 
    &=& \hat{P}_T T^{(\pm)}(x^5,x^6) \hat{P}_T, \\
  T^{\prime(\pm)}(x^5+2\pi R_5,x^6) 
    &=& -\hat{P}_T T^{\prime(\pm)}(x^5,x^6) \hat{P}_T, \\
  F^{(\pm)}(x^5+2\pi R_5,x^6) 
    &=& F^{(\pm)}(x^5,x^6) \hat{P}_T^{-1}, \\
  F^{\prime(\pm)}(x^5+2\pi R_5,x^6) 
    &=& -F^{\prime(\pm)}(x^5,x^6) \hat{P}_T^{-1}, \\
  N^{(\pm)}(x^5+2\pi R_5,x^6) &=& N^{(\pm)}(x^5,x^6),
\end{eqnarray}
where we have used matrix notation; $\hat{P}_T$ is given by 
$\hat{P}_T = {\rm diag}(1, 1, 1, -1, -1)$, which is obtained by 
projecting the matrix $P_T$ on the $SU(5)$ subspace.
With these boundary conditions, the correct low energy matter 
content follows.  Specifically, we find that $\{U, E\}, Q, D, L$ 
and $N$ come from $T^{(+)}, T^{\prime(+)}, F^{(+)}, F^{\prime(+)}$ 
and $N^{(+)}$, respectively.

The Yukawa couplings and the $U(1)_X$ breaking can be located 
either on the $(x^5, x^6) = (0, 0)$ or $(\pi R, 0)$ fixed point.
We here put them on the $(x^5, x^6) = (0, 0)$ fixed point as
\begin{eqnarray}
  {\mathcal L}_6 &=& \delta(x^5) \delta(x^6) 
    \int d^2\theta \Biggl\{ 
        (y_{T,1} T T H + 
        y_{T,2} T T' H + 
        y_{T,3} T' T' H)
\nonumber\\
&& +    (y_{F,1} T F \bar{H} +
        y_{F,2} T F' \bar{H} +
        y_{F,3} T' F \bar{H} +
        y_{F,4} T' F' \bar{H})
   +    (y_{N,1} F N H +
        y_{N,2} F' N H)
\nonumber\\
&& + Y (X \bar{X} - M_X^2) + \bar{X} N^2
        \Biggr\} + {\rm h.c.},
\label{eq:yukawa-5dmatter}
\end{eqnarray}
where we have omitted the superscript $(+)$ from each superfield 
$T^{(+)}, T^{\prime(+)}, F^{(+)}, F^{\prime(+)}$ and $N^{(+)}$; 
$H({\bf 5}, -2)$ and $\bar{H}({\bf 5}^*, 2)$ are components 
of the $\Phi$ field whose wavefunctions are nonvanishing at 
$(x^5,x^6) = (0,0)$, and $X({\bf 1}, 10), \bar{X}({\bf 1}, -10)$ 
and $Y({\bf 1}, 0)$ are chiral superfields localized on the 
$(x^5,x^6) = (0,0)$ fixed point.

How about $R$ symmetry and supersymmetry breaking?  We can impose 
$Z_{4,R}$ symmetry on the model as in sub-section \ref{subsec:su6-R}.
The $Z_{4,R}$ charge for the $M^{(+)}$ and $M^{(-)}$ chiral superfields 
are $0$ and $2$, respectively.  Then, all the couplings in 
Eq.~(\ref{eq:yukawa-5dmatter}) are allowed, while dangerous operators 
such as tree-level $d=5$ proton decay operators and a Higgs mass term 
are not.  As for the supersymmetry breaking, it can be either 
on the $(x^5, x^6) = (0, \pi R)$ or $(\pi R, \pi R)$ fixed point.
In either case, there is no specific relation for the three gaugino 
masses, $m_{SU(3)_C}, m_{SU(2)_L}$ and $m_{U(1)_Y}$, at low energies.

We finally comment on the phenomenology of the model with matter on the 
fixed line.  In this case, the quarks and leptons which would be unified 
into a single multiplet in the usual 4D grand unified theories come from 
different $SU(5)$ multiplets.  Specifically, $D$ and $L$ 
($Q$ and $U,E$) come from different (hyper)multiplets.
Therefore, proton decay from broken gauge boson exchange is absent
in this case \cite{Hall:2001pg}.  Furthermore, there is no unwanted 
$SU(5)$ relation among the low energy Yukawa couplings arising from 
the interactions given in Eq.~(\ref{eq:yukawa-5dmatter}) \cite{Hall:2001pg}.
This is reminiscent of the situation in certain string motivated
theories \cite{Witten:1985xc}.  Nevertheless, the theory still 
keeps the desired features of the usual 4D grand unified theory: 
the quantization of hypercharge and the unification of the 
three gauge couplings \cite{Witten:1985xc, Hall:2001pg}.
Therefore, this type of theory, with matter in the bulk, preserves 
(experimentally) desired features of 4D grand unified theories, 
while not necessarily having the problematic features, such as 
proton decay and fermion mass relations.

\subsection{A Theory of Flavor}
\label{subsec:hybrid}

In this sub-section, we present a model where some matter lives on a 
fixed point and some on a fixed line.  An important point 
of this model is that it partially explains the mass hierarchies among 
the three generations of quarks and leptons, and simultaneously explains 
why the $SU(5)$ mass relation holds for the heavier generation 
but fails for lighter generations.  The mechanism presented here 
also applies for the 5D $SU(5)$ model discussed in 
Refs.~\cite{Kawamura:2001ev, Hall:2001pg}. 

We put one generation of quarks and leptons on the $(x^5, x^6) = (0, 0)$ 
fixed point.  Since this fixed point preserves $SU(5) \otimes U(1)_X$ 
symmetry, we introduce $T_3({\bf 10}, 1)$, $\bar{F}_3({\bf 5}^*, -3)$ 
and $N_3({\bf 1}, 5)$ chiral superfields.  The meaning of the suffix 
$3$ becomes apparent later when we identify these fields as the third 
generation quarks and leptons.  The other two generations 
are located on the fixed line $x^6 = 0$.  Thus, we introduce 
hypermultiplets ${\cal T}_i({\bf 10}, 1)$, ${\cal T}'_i({\bf 10}, 1)$, 
${\cal F}_i({\bf 5}^*, -3)$, ${\cal F}'_i({\bf 5}^*, -3)$ and 
${\cal N}_i({\bf 1}, 5)$ on this line, where $i=1,2$ represents 
the generation index.  The $Z_{4,R}$ symmetry and supersymmetry breaking 
are the same as before.  Since the first two generations are located 
on the fixed line, proton decay is suppressed.  The supersymmetry 
breaking occurs either at $(x^5, x^6) = (0, \pi R)$ or $(\pi R, \pi R)$ 
fixed point and is mediated by gaugino interaction.

A distinctive feature of the present setup comes from the structure 
of the Yukawa couplings.  They are located on the $(x^5, x^6) = (0, 0)$
fixed point.  We allow all the couplings of the forms $T T H$, 
$T F \bar{H}$ and $F N H$, where $T$ collectively represents 
$T_i^{(+)}$, $T_i^{\prime(+)}$ and $T_3$ and similarly for $F$ and $N$.
Specifically, we introduce 
\begin{eqnarray}
  {\mathcal L}_6 &=& \delta(x^5) \delta(x^6) 
    \int d^2\theta \Biggl\{ \Bigl(
        (y_T)_{33} T T H + (y_F)_{33} T F \bar{H} + (y_N)_{33} F N H \Bigr)
\nonumber\\
&& +    \sum_{i=1}^{2} \Bigl( 
        (y_{T,1})_{3i} T T_i H + (y_{T,1})_{i3} T_i T H + 
        (y_{T,2})_{3i} T T'_i H + (y_{T,2})_{i3} T'_i T H 
\nonumber\\
&& \qquad + (y_{F,1})_{3i} T F_i \bar{H} + (y_{F,1})_{i3} T_i F \bar{H} + 
        (y_{F,2})_{3i} T F'_i \bar{H} + (y_{F,2})_{i3} T'_i F \bar{H} 
\nonumber\\
&& \qquad + (y_{N,1})_{3i} F N_i H + (y_{N,1})_{i3} F_i N H + 
        (y_{N,2})_{i3} F'_i N H) \Bigr)
\nonumber\\
&& +    \sum_{i,j=1}^{2} \Bigl( 
        (y_{T,1})_{ij} T_i T_j H + 
        (y_{T,2})_{ij} T_i T'_j H + 
        (y_{T,3})_{ij} T'_i T'_j H
\nonumber\\
&& \qquad + (y_{F,1})_{ij} T_i F_j \bar{H} +
        (y_{F,2})_{ij} T_i F'_j \bar{H} +
        (y_{F,3})_{ij} T'_i F_j \bar{H} +
        (y_{F,4})_{ij} T'_i F'_j \bar{H}
\nonumber\\
&& \qquad + (y_{N,1})_{ij} F_i N_j H +
        (y_{N,2})_{ij} F'_i N_j H \Bigr)
\nonumber\\
&& + Y (X \bar{X} - M_X^2) + \bar{X} N^2
        \Biggr\} + {\rm h.c.},
\end{eqnarray}
where we have dropped the superscript $(+)$ from the fields which come 
from bulk hypermultiplets, and omitted order one coefficients in units 
of fundamental scale $M_*$ of the theory. Yukawa couplings between 
members of the lighter two generations are also located at 
$(x^5, x^6) = (\pi R, 0)$.

An important point is that since the couplings of bulk fields are 
suppressed by the volume of the extra dimension, we obtain the 
Yukawa structure that the couplings for the first two generations 
are suppressed compared with the third generation ones.
Specifically, we obtain the following structure for the Yukawa couplings 
\begin{equation}
  y_u \sim y_d \sim y_e \sim y_{\nu} \sim
    \pmatrix{
        \epsilon^2 & \epsilon^2 & \epsilon \cr
        \epsilon^2 & \epsilon^2 & \epsilon \cr
        \epsilon   & \epsilon   & 1        \cr
    },
\label{eq:ferm-yukawa}
\end{equation}
where $\epsilon = 1/(\pi R M_*)$ is a small parameter representing 
the volume suppression factor.  This provides a partial understanding of 
the mass hierarchy among the generations from a geometrical viewpoint.
In particular, it is very natural to identify the fields on the fixed 
line with the first two generations of  matter and those on the fixed point 
with the third generation, since the former receive wavefunction 
suppressions while the latter does not.

Another important point is that since the $(x^5, x^6) = (0, 0)$ 
fixed point respects $SU(5) \otimes U(1)_X$ gauge symmetry, the Yukawa 
couplings among the fields localized on this fixed point must 
respect $SU(5)$ relations.  Since the fields on this point are 
identified with the third generation matter, we obtain the relation 
$(y_d)_{33} = (y_e)_{33}$, which means that the $b/\tau$ unification 
is preserved in the present model.  On the other hand, the first two 
generations of matter come from the hypermultiplets located on the 
fixed line $x^6 = 0$ so that they do not respect the $SU(5)$ mass 
relations.  This means that we do not get unwanted fermion mass 
relations such as $m_s/m_d = m_\mu/m_e$.  Together with the argument 
of volume suppression leading to Eq.~(\ref{eq:ferm-yukawa}), 
the present setup provides an understanding for why the heaviest 
generation respect $SU(5)$ mass relation while the lighter ones 
do not.

\section{6D $SU(6)$ Unified Model on $T^2/Z_4$}
\label{sec:t2z4}

One less than ideal feature of the $T^2/(Z_2 \times Z_2')$ orbifold
model discussed in the previous section is the non-universal power-law
running of the gauge couplings above the compactification scale.  
As discussed in sub-section \ref{subsec:su6running}, this running
leads to corrections to the standard supersymmetric $SU(5)$ prediction 
of $\sin^2\theta_w$.  These corrections can be suppressed 
by taking $M_*$ to be not far above $M_c$.  Here we instead
consider an alternative $SU(6)$ model in which the running of the gauge
couplings is just as in the MSSM, even above the compactification scale.

The orbifold for this model is $T^2/Z_4$.  As before the only bulk
fields are those of a 6D $N=2$ vector multiplet.  Defining 
$z \equiv x_5+ix_6$ and $\partial \equiv \partial_5 - i\partial_6$, 
the bulk action may be written using 4D $N=1$ language as 
\cite{Arkani-Hamed:2001tb}   
\begin{eqnarray}
  S &=& \int d^6 x \Biggl\{
  {\rm Tr} \Biggl[ \int d^2\theta \left( \frac{1}{4 k g^2} 
  {\cal W}^\alpha {\cal W}_\alpha + \frac{1}{k g^2} 
  \left( \Phi^c \partial \Phi   - \frac{1}{\sqrt{2}} \Sigma 
  [\Phi, \Phi^c] \right) \right) + {\rm h.c.} \Biggr] 
\nonumber\\
  && + \int d^4\theta \frac{1}{k g^2} {\rm Tr} \Biggl[ 
  (\sqrt{2} \partial^\dagger + \Sigma^\dagger) e^{-V} 
  (-\sqrt{2} \partial + \Sigma) e^{V}
  + \Phi^\dagger e^{-V} \Phi  e^{V}
  + {\Phi^c}^\dagger e^{-V} \Phi^c e^{V} 
\Biggr] \Biggr\},
\label{eq:t2z4action}
\end{eqnarray}
in the Wess-Zumino gauge.  Here the gauge field components $A_5$ and 
$A_6$ are both contained in $\Sigma$, so that both $\Phi$ and $\Phi^c$
transform linearly under gauge transformations.  When expressed in terms
of components, this action and that of Eq.~(\ref{eq:5daction}) have
identical forms.  The orbifold of the present model will
preserve a different 4D $N$=1 supersymmetry than the orbifold of the
previous model (namely, one in which $A_5$ and $A_6$ are paired in the 
same superfield), and we have chosen to make
this different 4D $N$=1 supersymmetry manifest. 

The orbifold $T^2/Z_4$ is constructed by identifying points of the
infinite plane $R^2$ under three operations,  
${\cal Z}: z \rightarrow i z$, ${\cal T}_5: 
z \rightarrow z+2\pi R$ and ${\cal T}_6: 
z \rightarrow z+2\pi i R$.  The identifications for the fields under
${\cal Z}$ are taken to be 
\begin{eqnarray}
  V(i z) &=& P_Z V(z) P_Z^{-1},
\\
  \Sigma(i z) &=& -i P_Z \Sigma(z) P_Z^{-1},
\\
  \Phi(i z) &=& -P_Z \Phi(z) P_Z^{-1},
\\
  \Phi^c(i z) &=& -i P_Z \Phi^c(z) P_Z^{-1},
\end{eqnarray}
and the identifications under ${\cal T}_5$ and ${\cal T}_6$ are
\begin{eqnarray}
  V(z+2\pi R) &=& P_T V(z) P_T^{-1},
\\
  \Sigma (z+2\pi R) &=& P_T \Sigma (z) P_T^{-1},
\\
  \Phi(z+2\pi R) &=& P_T \Phi (z) P_T^{-1},
\\
  \Phi^c(z+2\pi R) &=& P_T \Phi^c(z) P_T^{-1},
\end{eqnarray}
and 
\begin{eqnarray}
  V(z+2\pi i R) &=& P_T V(z) P_T^{-1},
\\
  \Sigma (z+2\pi i R) &=& P_T \Sigma (z) P_T^{-1},
\\
  \Phi(z+2\pi i R) &=& P_T \Phi (z) P_T^{-1},
\\
  \Phi^c(z+2\pi i R) &=& P_T \Phi^c(z) P_T^{-1},
\end{eqnarray}
respectively.  As before, we take
\begin{eqnarray}
  P_Z &=& {\rm diag}(1, 1, 1, 1, 1, -1),
\\
  P_T &=& {\rm diag}(1, 1, 1, -1, -1, -1),
\end{eqnarray}
which breaks the $SU(6)$ gauge group to $SU(3)_C \otimes SU(2)_L
\otimes U(1)_Y \otimes U(1)_X$ at low energies.  It is not difficult
to see that the only massless zero modes besides those of $V$
come from $\Phi$ and have the quantum numbers of the two MSSM 
Higgs doublets.

This orbifold has two fixed points located at $(x_5, x_5)=(0,0)$ and 
$(\pi R, \pi R)$.  The remaining supersymmetry on both of these fixed
points is 4D $N$=1, and the gauge symmetries are $SU(5)\otimes U(1)_X$
and $SU(3)_C \otimes SU(2)_L \otimes U(1)_Y \otimes U(1)_X$,
respectively.  The important distinction between this orbifold and the
$T^2/(Z_2 \times Z_2')$ orbifold considered previously is that there
are no longer 5D fixed lines on which only 4D $N$=2 supersymmetry
remains.  Thus, the bulk 4D $N$=4 supersymmetry ensures that the terms
quadratically {\it and} linearly sensitive to the cutoff vanish in
Eq.~(\ref{eq:powerlaw}).  In fact, explicit calculation of the KK
modes for $V$, $\Sigma$, $\Phi$, and $\Phi^c$ reveals that at {\it each}
massive level the states are arranged in $N$=4 multiplets, so that the only
states that contribute to the running of gauge couplings are the massless
zero modes.  This means that Eq.~(\ref{eq:rge}) holds with each of the
$\gamma$ coefficients set to zero.  The fundamental scale $M_*$ should
then be chosen to be close to the usual 4D unification scale $M_U\simeq
2 \times 10^{16}$ GeV.  The compactification scale should not be more
than a factor $\sim 10$ lower, so that the the theory remains
perturbative up to $M_*$, but the prediction for $\sin^2\theta_w$ 
is insensitive to the precise value of $M_c/M_*$.  

If the quarks and leptons are placed on the $SU(5)\otimes
U(1)_X$ preserving fixed point, $SU(5)$ mass relations follow,
and the setups of sub-sections \ref{subsec:5dfermion} and
\ref{subsec:hybrid} cannot be used to alter these relations because
of the absence of fixed lines with reduced supersymmetry.  Thus we
choose instead to introduce matter on the $SU(3)_C \otimes SU(2)_L 
\otimes U(1)_Y \otimes U(1)_X$ preserving fixed point.  The
discussions of $U(1)_X$ breaking, $R$ symmetry, and supersymmetry
breaking from sub-sections 
\ref{subsec:su6u1x} -- \ref{subsec:su6susybreak} then carry over 
in the obvious way.

\section{5D $SU(6)$ Unified Model on $S^1/Z_2$}
\label{sec:5d-su6}

In this section, we give a model in which Higgs fields arise 
from the extra dimensional components of the gauge fields.
The model is based on a 5D $N=1$ supersymmetric gauge theory with 
the extra dimension $y$ compactified on $S^1/Z_2$.
We take the gauge group to be $SU(6)$.  In terms of 4D $N=1$ 
supersymmetry language, we have a vector superfield $V$ and a chiral 
superfield $\Phi$ both in the adjoint representation of the $SU(6)$.

The orbifold is defined by specifying boundary conditions for the 
bulk fields under two operations ${\cal Z}: y \rightarrow -y$ and 
${\cal T}: y \rightarrow y+2\pi R$ as 
\begin{eqnarray}
  V(-y) &=& P_Z V(y) P_Z^{-1}, \\
  \Phi(-y) &=& -P_Z \Phi(y) P_Z^{-1},
\end{eqnarray}
and
\begin{eqnarray}
  V(y+2\pi R) &=& P_T V(y) P_T^{-1}, \\
  \Phi(y+2\pi R) &=& P_T \Phi(y) P_T^{-1},
\end{eqnarray}
respectively.  Here, $V = V^A T^A$, $\Phi = \Phi^A T^A$, 
$P_Z$ and $P_T$ are $6 \times 6$ matrices.

We take $P_Z={\rm diag}(1,1,1,1,1,-1)$ and $P_T={\rm diag}(1,1,1,-1,-1,-1)$, 
so that $SU(3)_C \otimes SU(2)_L \otimes U(1)_Y \otimes U(1)_X$ 
remains unbroken at low energies.  Then, the transformation properties 
for the bulk fields are explicitly given in 
Eqs.~(\ref{eq:V6trans}, \ref{eq:phi6trans}) with the first and second 
signs representing quantum numbers under ${\cal Z}$ and ${\cal T}$ 
respectively.  We find that two zero-mode chiral superfields
from $\Phi$, which have $(+,+)$ transformation properties, have 
precisely the quantum numbers of the two Higgs doublets under the 
standard model gauge group.  Therefore, we identify these massless 
fields as the MSSM Higgs fields.  In contrast to the model presented 
in previous sections, here some components of the Higgs doublets are 
extra dimensional components of the heavy unified gauge bosons.

The quarks and leptons are introduced either on $y=0$ or $y = \pi R$ 
fixed point.  Here, for an illustrative purpose, we introduce them 
on the $y=0$ fixed point where $SU(5) \otimes U(1)_X$ gauge invariance 
is manifest.  The $y=\pi R$ fixed point case can be worked 
out quite similarly.  Since the $y=0$ fixed point preserves 
only $SU(5) \otimes U(1)_X$ gauge symmetry, we can introduce 
three generations of quarks and leptons, $3 \times [T({\bf 10}, 1),$ 
$F({\bf 5}^*, -3),$ $N({\bf 1}, 5)]$.  Here, we have normalized 
the $U(1)_X$ charges to match the convention, $U(1)_X = SO(10)/SU(5)$.

An immediate difficulty in the present model compared with the 
previous models is that, since the Higgs fields 
are the extra dimensional components of the gauge fields, 5D gauge 
invariance prevents us from introducing Yukawa couplings between 
the Higgs field $\Phi$ and the quarks and leptons.
Specifically, the Yukawa couplings are forbidden by non-linear 
transformation of the $\Phi$ field, $\Phi \rightarrow e^\Lambda
(\Phi - \sqrt{2} \partial_y) e^{-\Lambda}$, under the 5D gauge 
transformation.  To circumvent this problem, we consider the Wilson 
line operator, ${\cal P} \exp(\int_0^{2\pi R} \Phi dy)$, where 
${\cal P}$ represents the path ordered product.  This object transforms 
linearly under the gauge transformation, so that we can couple it to the 
quark and lepton fields.  Since the $y=0$ fixed point preserves 
only $SU(5) \otimes U(1)_X$ gauge symmetry, we consider the subsets 
of the $6 \times 6$ Wilson line matrix which contain linear terms 
in the zero modes.  They are given by the upper-right five-dimensional 
column vector $H$ and the lower-left five-dimensional row vector 
$\bar{H}$:
\begin{eqnarray}
  H(x^\mu) &=& 
    \left. {\cal P} \exp \left(\int_0^{2\pi R} \Phi dy \right) 
    \right|_{({\bf 5}, -2)_{y = 2\pi R},\, ({\bf 1}, -10)_{y = 0}},
\\
  \bar{H}(x^\mu) &=& 
    \left. {\cal P} \exp \left(\int_0^{2\pi R} \Phi dy \right) 
    \right|_{({\bf 1}, 10)_{y = 2\pi R},\, ({\bf 5}^*, 2)_{y = 0}}.
\end{eqnarray}
Here, $H$ transforms as $({\bf 5}, -2)$ under the gauge transformation 
$SU(5) \otimes U(1)_X$ at $y=2\pi R$ but as $({\bf 1}, -10)$ 
under that at $y=0$.  Similarly, $\bar{H}$ transforms as 
$({\bf 1}, 10)$ at $y=2\pi R$ and as $({\bf 5}^*, 2)$ at $y=0$.

Introducing the brane fields $X({\bf 1}, 10)$ and $\bar{X}({\bf 1}, -10)$,
we can write down gauge-invariant non-local interactions among 
the Higgs field and quarks and leptons,
\begin{equation}
  S = \int d^4 x \int d^2\theta \left\{
    y_T (T T)|_{y =2\pi R} X|_{y =0} H
  + y_F (T F)|_{y =0} \bar{X}|_{y =2\pi R} \bar{H}
  + y_N (F N)|_{y =2\pi R} X|_{y =0} H \right\},
\label{eq:ap-nonlocal}
\end{equation}
with each term suppressed by appropriate powers of the fundamental 
scale $M_*$.  Here we will not consider the physics which may generate 
these non-local operators, but simply take a viewpoint that they can 
be written in the effective field theory since they are gauge invariant.
We also introduce the brane-localized superpotential 
\begin{equation}
  S = \int d^4 x dy\, \delta(y) \int d^2\theta 
    \left\{ Y (X \bar{X} - M_X^2) + \bar{X} N^2 \right\},
\label{eq:ap-local}
\end{equation}
where $Y$ is the singlet superfield.  This superpotential forces 
$X$ and $\bar{X}$ to have vacuum expectation values 
$\langle X \rangle = \langle \bar{X} \rangle = M_X$, breaking
$SU(3)_C \otimes SU(2)_L \otimes U(1)_Y \otimes U(1)_X$ down to 
the standard model gauge group.  Then, assuming $M_X \sim M_*$, 
Eq.~(\ref{eq:ap-nonlocal}) gives the usual Yukawa couplings 
at low energies.  The Majorana masses for the right-handed neutrinos 
of order $M_X$ also arise from Eq.~(\ref{eq:ap-local}), so that 
small neutrino masses are generated through the see-saw mechanism.

We finally comment on the phenomenological issues.
The present model does not suffer from the power-law correction for 
$\sin^2\theta_w$; above the compactification scale $1/R \approx M_U$, 
differential running between the three standard-model gauge couplings 
is logarithmic.  Therefore, the situation is similar to the $SU(5)$ 
case discussed in Ref.~\cite{Hall:2001pg}, and retains an exciting 
possibility that the proton decay caused by dimension 6 operators 
may be seen in the near future. (This proton decay is absent if we 
put quarks and leptons on the $y= \pi R$ fixed point, since the 
wavefunctions for the $X, Y$ gauge bosons vanish there.)

Proton decay from dimension 5 operators is forbidden in a similar way 
to the $SU(5)$ case in Ref.~\cite{Hall:2001pg}, by imposing a $U(1)_R$ 
symmetry.  The higher KK modes of $\Phi$ form mass terms together 
with $V$, becoming a part of the massive vector multiplets.  
The $U(1)_R$ charge assignment is given as $\Phi(0), T(1), F(1), 
N(1), X(0), \bar{X}(0), Y(2)$.  A good thing here compared with the 
$SU(5)$ case is that the $U(1)_R$ charges for the Higgses, $\Phi$, are 
automatically fixed to the desired value: the bulk $U(1)_R$ is just a 
subgroup of the $SU(2)_R$ automorphism group of $N=2$ supersymmetry 
algebra.  This $U(1)_R$ also forbids dimension 4 proton decay, since 
it contains $R$-parity as a subgroup.

As for the $\mu$ term and the supersymmetry breaking, the situation is 
the same with the $SU(5)$ case. If the supersymmetry breaking occurs on 
the $y = \pi R$ fixed point, gaugino mediation could naturally occur 
and there is no supersymmetric flavor problem.  Realistic fermion masses 
could result from mixing of brane and bulk matter \cite{Hall:2001pg}.
(In the case of matter on the $y = \pi R$ fixed point, realistic 
fermion masses are more easily obtained, since the Yukawa couplings 
need only preserve $SU(3)_C \otimes SU(2)_L \otimes U(1)_Y \otimes 
U(1)_X$ on that fixed point.)

\section{Conclusions}

In this paper we have explored the idea that the Higgs doublets of the
minimal supersymmetric standard model originate from the same higher
dimensional supermultiplet that contains the standard model gauge fields.
This requires an extension of the standard model gauge group, and 
is particularly suited to the situation where gauge symmetry breaking
is induced by orbifold boundary conditions. The Higgs doublets emerge 
naturally as zero modes, and despite the bulk gauge symmetry, there
are orbifold fixed points where independent Yukawa couplings of the
Higgs to matter are allowed. This idea demonstrates that it is natural 
for a supersymmetric gauge theory in higher dimensions to give
states below the fundamental scale which are vector-like with respect 
to the gauge group. 

We have written several explicit models with such a
unification of the Higgs doublets with the standard model gauge
bosons. In the minimal case, where the electroweak gauge group is
extended to $SU(3)$, the compactification scale could be as low as a
few TeV. Alternatively the compactification scale could be at the
unification scale, and we have shown that the appearance of Higgs
doublets from compactification of the gauge multiplet allows a
preservation of the successful weak mixing angle prediction.

In the supersymmetric standard model some mechanism is needed to solve
the $\mu$ problem --- that is to explain why the vector-like Higgs
doublets only pick up a mass at the scale of supersymmetry
breaking. Similarly, a mechanism is needed to break supersymmetry in a
way which does not introduce too large flavor violations.  
Such mechanisms are also required in the present case, and, given the
higher dimensional orbifold context, a particularly natural possibility
emerges. An $R$ symmetry protects the $\mu$ parameter, and this is
broken on the same brane where supersymmetry is broken, yielding a TeV
mass for the Higgs. Furthermore, as long as this brane is not the one
containing matter, the supersymmetry breaking preserves flavor by the
gaugino mediation mechanism.  The higher dimensional framework not
only provides a reason for the existence of the Higgs, but allows a
simple explanation for why it is light.

\section*{Acknowledgments}

Y.N. thanks the Miller Institute for Basic Research in Science 
for financial support.  This work was supported by 
the Department of Energy under contract DE-AC03-76SF00098 
and the National Science Foundation under contract PHY-95-14797.

\end{document}